# Collective neutral excitations as sensitive probe for the quality of 2D charge carrier systems in ultra-pure GaAs quantum wells

Ursula Wurstbauer[1,2*], Michael J. Manfra[3,4], Ken W. West[5], and Loren N. Pfeiffer[5]

[1]Institute of Physics, University of Münster, Münster, Germany
[2]Center for Soft Nanoscience (SoN), University of Münster, Münster, Germany
[3]Department of Physics and Astronomy, School of Materials Engineering, and School of Electrical and Computer Engineering, Purdue University, West Lafayette, Indiana, USA
[4] Microsoft Quantum Lab West Lafayette, West Lafayette, IN, 47907 USA.
[5]Department of Electrical Engineering, Princeton University, Princeton, NJ, USA

*corresponding author: wurstbauer@uni-muenster.de

**Abstract**

Ultra-clean low-dimensional interacting charge carrier systems are the basis to explore correlated states and phases. We report the observation of very narrow collective intersubband excitations (ISBE) of 2D electron systems (2DESs) with ultra-high mobilities in high quality GaAs quantum well structures. These findings from resonant inelastic light scattering (RILS) experiments are used as tools for exploration of links between transport mobility and collective electron behavior in 2DES of high perfection. We find that the linewidths of collective ISBE modes can be very narrow with values smaller than 80 μeV. Comparison of ISBE measurements from several high-mobility samples exhibits a variation in linewidth of more than a factor of two. There is, however, a surprising lack of direct correlation between ISBE linewidth with mobility in the range $15 \cdot 10^6 \, cm^2/Vs \leq \mu \leq 24 \cdot 10^6 \, cm^2/Vs$. ISBE by RILS are discussed as a sensitive probe to characterize the interacting electron systems for fractional quantum Hall effect (FQHE) studies.

**Highlights:**

- Observation of very narrow intersubband excitations of high mobility ($\mu \geq 15 \cdot 10^6 \, cm^2/Vs$) 2DES.
- Single particle-like excitation with spin-flip indicative for spin-orbit effects.
- Plasmon-like charge density excitation sensitive to homogeneity and purity.
- RILS on ISBE powerful probe to characterize 2DES at relaxed experimental conditions.

**Introduction**

Resonant Inelastic Light Scattering (RILS) on collective electronic excitations has been established as a powerful method to study interacting charge carrier systems in three-, two-, one- and zero-dimensional systems as well as in correlated phases such as quantum and fractional quantum Hall states [1]–[11]. Inelastic scattering of light on media has been predicted in 1921 by Smekal [12] and experimentally confirmed for various liquids in 1928 by Raman and Krishan [13]. Inelastic scattering of light by gases, liquids and solids was inspired by the Compton effect [14],[15],[16] due to the conservation principles and therefore referred to as "optical analogue of the Compton effect" even though being aware that quantum mechanical effects at the scattered media seems to play a role [13],[17]. Already in his Nobel lecture written end of the year 1930, Sir Raman stated that "the universality of the phenomenon, the convenience of the experimental technique and the simplicity of the spectra obtained enable the effect to be used as an experimental aid to the solution of a wide range of problems in physics and chemistry" [17].

By considering only solid-state based systems, this generic prediction greatly surpassed the expectation since Raman spectroscopy and microscopy is commonly used as a fast, versatile and non-invasive method in (semiconductor) industry as well as for fundamental research and material science on bulk and on low-dimensional quantum- and nano-systems. Raman spectroscopy and microscopy can even be applied in-operando and under changing environments. Raman scattering on collective lattice excitations provides access to a plethora of properties of condensed matter comprising but not limited to lattice symmetry, material composition, strain, stress, temperature, estimation of defect and charge carrier density [3],[18]–[20]. For this reason, the unique vibrational properties are also called "phonon fingerprint" of the crystals. In addition to information on the lattice degree of freedom, electronic properties are accessible under resonant excitation [3],[21]–[23]. Coupled excitations can impact the materials properties itself [24],[25]. In recent years, Raman spectroscopy is again on the rise in the intense and still growing interest on two-dimensional crystals and their hetero- and hybrid-structures [26],[27].

Inelastic scattering of light on collective electronic excitations is observable under resonant excitations. RILS, a.k.a. electronic Raman scattering, on neutral elementary excitations allows to study interacting electron systems and quasiparticle excitations in emergent quantum states particularly in doped semiconductor heterostructures. Seminal results include the experimental observation of collective electronic intersubband excitation (ISBE) in modulation doped GaAs/AlGaAs quantum wells (QW) by Pinczuk *et al.* [11] and nearly simultaneously in a quasi-two-dimensional electron system hosted in a GaAs/AlGaAs heterojunction by Abstreiter *et al.* [28]. Hereby, spin and charge-density excitation for both, intra- and intersubband excitations can be clearly distinguished by their distinct polarization selection rules. These excitations have been observed for quasi two-dimensional electron and hole systems also in the quantum Hall regime [6],[7],[29],[30],[31],[6],[7],[32],[33] as well as for low-dimensional structures such as quantum dots and nanowires [4],[34],[35] [15] including ultra-thin core-shell nanowires [10] and, more recently, also between moiré-minibands in twisted two-dimensional semiconducting crystals [36].

In continued pioneering efforts Aron Pinczuk applied RILS to emergent quantum phases mainly supported by quasi two-dimensional electron system hosted in ultrapure GaAs/AlGaAs heterostructures such as single and tunnel coupled QWs [1],[8],[37]–[41]. In a groundbreaking work Pinczuk *et al.* experimentally observed for the first time the theoretically predicted neutral charge and spin density excitations of the quantum liquid at the fractional quantum Hall effect (FQHE) state at filling factor 1/3 [1]. By pushing the experimental boundaries and advancing the RILS method to temperatures of only a few hundred Millikelvin - the ultimate lower limit at that time - and by applying large magnetic fields, they observed a series of RILS modes of the 1/3 FQHE state at the critical points in the respective wave vector dispersion: the so-called roton minimum at finite momentum, the gap excitation at the long-wave-length limit and a spin-excitation at the bare Zeeman energy according to Lamor's theorem [1].

With continuously improved conditions in the molecular beam epitaxial growth of the heterostructure including optimization of heterostructure design, the purity and consequently the mobilities of the two-dimensional charge carrier systems significantly increased [42]–[44]. With improved quality of the heterostructures, an increasing number of fractional quantum Hall states were observed. Among those FQHE states are some with non-abelian ground states such as the 5/2 state in the second LL. These exotic states are observed in conjunction with charge ordered phases and Wigner solids that can result in competing quantum states [45]–[48]. The low-lying collective excitations and their dispersion relations govern the rich correlation physics of those quantum phases. Studying these neutral excitations and their polarization properties by RILS is till now a very powerful tool to uncover the nature of their ground states and to probe potential spin polarization and competition of nearly degenerate phases [49],[37],[39],[41]. Based on circularly polarized RILS experiments on FQH liquids in the lowest LL described by flux-two composite fermions [46], strong evidence for the existence chiral spin-2 graviton modes as condensed matter analogues of gravitons have been recently reported demonstrating the relevance of quantum geometry [8].

The enhanced mobility of the two-dimensional charge carrier systems and purity of the hosting GaAs/AlGaAs crystals over the years stimulated the research on interaction-driven physics and emergent quantum phases. The field has substantially advance since the early days. This is an ongoing process and further optimization of the quality of the charge carrier systems requires advanced probes that are on the one-hand side sensitive to electron correlation and symmetry of the underlying quantum wells and on the other-hand side provide fast, easy and non-destructive access without the need of further processing e.g. in micro- or nanoscale circuitries. RILS is one of the very few methods that allow direct access to low-lying neutral excitation of interacting electron systems and is rather fast and versatile due to its far-field optical nature.

Here, we demonstrate that the line-width of the ISBE charge density excitation (CDE) in RILS spectra taken at zero-field at cryogenic temperatures of about 4.5K is a sensitive probe for the quality of the charge carrier system for FQHE studies that can be more sensitive than Hall mobilities measured in van-der Pauw geometry. Moreover, we demonstrate evidence that the line-shape of the so called single-particle excitation band in depolarization scattering geometry in RILS spectra of ISBE is indicative for zero-field splitting of the conduction band due to Rashba- and/or Dresselhaus spin-orbit interaction [50] similar as it has been demonstrated for intraband excitations [33],[31]. Together with scanning photoluminescence spectroscopy allowing access to local fluctuations in the charge carrier density, scanning RILS measurements at rather relaxed experimental conditions at zero-field and liquid helium temperature of about 4.5K provide fast and comprehensive characterization of the quality of the interacting electron system that is of paramount importance for the formation and stabilization of new correlated states of matter.

**Main**

All sample investigated in this study consist of a high-mobility quasi two-dimensional electron system confined in single GaAs QW with AlGaAs barriers that are symmetrically modulation-doped with silicon delta layers from both sides as schematically depicted in Figure 1(a) if not explicitly stated otherwise. The lowest conduction band (CB) at the zone-center (Γ-point) together with the energetic position of the Fermi-energy $E_F$ both simulated using *nextnano3* software [51] is included in Figure 1(a). The QW widths of the different samples varies between 23nm and 31nm. The ultra-clean structures have been grown by molecular beam epitaxy either on polar (001)-GaAs or on nonpolar (110)-GaAs substrates with optimized growth conditions for each surface. Small pieces of the grown wafer with a width of about 5mm and a length of about 25mm have been mounted on the cold finger of an optical bath-cryostat cooled with liquid $^4$He and operated at a temperature of 4.5K. The cryostat is equipped with windows for direct optical free-space access for measurements without applied magnetic field. For measurements in the quantum Hall regime with an applied magnetic field the sample were mounted under on tilt of about 20° between sample normal and the magnetic field direction on the cold finger of a $^3$He/$^4$He dilution refrigerator with a base temperature of about 42mK also equipped with windows. The cold finger is inserted in the bore of a 16T superconducting electromagnet. Further details on the setup used for magnetospectroscopy can be found elsewhere [41].

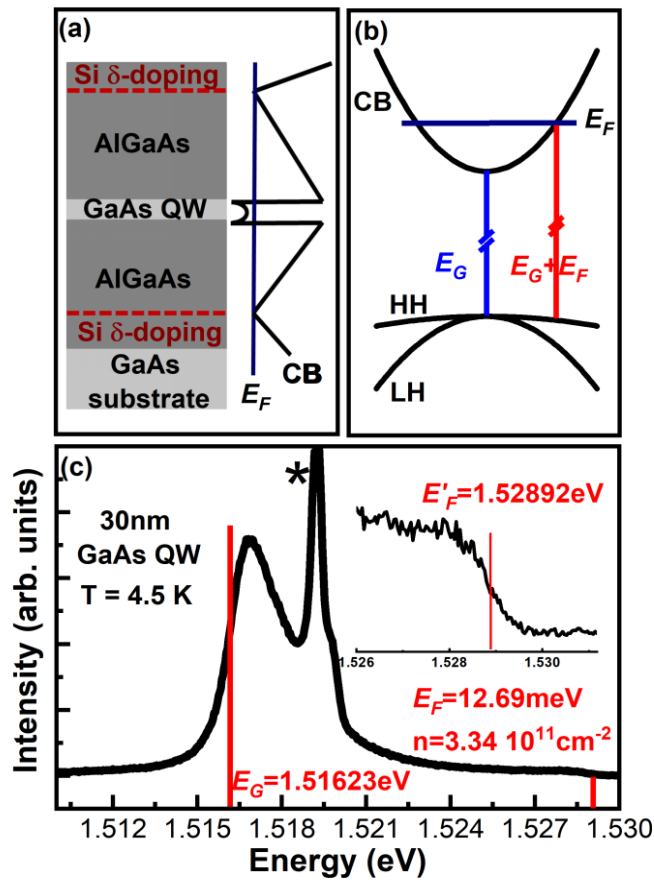

*Figure 1: (a) Scheme of the layer stack of the ultra-pure MBE-grown modulation doped GaAs/AlGaAs heterostructure hosting the high-mobile quasi two-dimensional charge carrier system with the lowest CB with the Fermi level $E_F$ at the $\Gamma$-point sketched next to it. (b) Scheme of the parabolic CB and VB with $E_F$ and the relevant energies to determine $E_F$ and hence the charge carrier density from PL measurements. (c) PL spectrum taken on a 30nm GaAs QW structure at 4.5K. The steep onset at the band gap energy $E_G$ and the Fermi-edge with the characteristic shape of the Fermi-Dirac distribution function magnified at the inset allows the determination of $E_F$. The sharp feature marked by an asterisk is assigned to a localized exciton donor bound exciton (DX-center).*

The charge carrier mobilities and densities from Hall and Shubnikov-de Hass measurements in van der Pauw geometry were determined on different pieces from the same wafers for all heterostructures in this study at $T$ = 300mK after low-temperature illumination. The charge carrier densities were independently measured by photoluminescence (PL) investigations at 4.5K. As schematically depicted in Figure 1(b), illumination above the band gap generate photoexcited electron-hole pairs that relax under emission of phonons to the lowest CB and the topmost valence band states inside the GaAs

QW. Due to Pauli blocking, the electrons can only relax to the Fermi-energy $E_F$. Weakly bound electron-hole pairs form and recombine under emission of light with typical wavelengths between band-gap onset $E_G$ and ($E_0+E_F$). In this way, the two-dimensional charge carrier density can be determined from a PL spectrum by calculating the density $n = E_F \cdot \frac{m^*}{\pi \hbar^2}$, with m* the effective electron mass of the GaAs CB band of $m^* = 0.065 m_0$ with $m_0$ being the free electron mass and $\hbar$ the reduced Planck constant [52]. At typical low-temperature (4.5K) PL spectrum taken on a 30nm GaAs QW grown on (001) surface and excited with a wavelength of about 800nm and a macroscopically spot size of about 300µm x 2mm using a cylindrical lens is shown in Figure 1 (c). The steep onset at the low energy side marks the gap energy $E_G$ that can be determined rather precisely taking the first derivative of the spectrum. On the high energy side of the spectrum, the line-shape at the Fermi-edge follows the Fermi-Dirac distribution function [see inset of Figure 1(c)]. Fitting the spectra with the distribution function allows the determination of $E_F$ and hence an estimation of the charge carrier density that is for most of the investigated samples in good agreement with the values observed from transport measurements. The sharp peak in the PL spectrum marked with an asterisk is a localized defect bound exciton (Coulomb-coupled electron-hole pair bound to a defect, most likely a DX center [53]). We would like to note that only sample were investigated for which the charge carrier density did not change in dependence of intensity and energy of the illuminating laser light in the relevant parameter range.

The aim of this study is to use RILS on collective electronic excitations as a sensitive probe for the quality of the heterostructures and their suitability for the investigation of correlation physics in the (fractional) quantum Hall regime. The relevant types of collective excitations of quasi-two-dimensional electron systems (2DES) hosted in QWs or heterojunctions are depicted in Figure 2(a). Without magnetic field we differentiate between intra- and intersubband excitations [4]. Each support spin conserving excitations and ones with spin flip. In the case of a spin-unpolarized 2D charge carrier systems in a QW with parabolic bands, the two spin-subsystems oscillated in phase for a spin-conserving excitation and with a $\pi$-phase shift for excitations with spin flip. While the charge density excitations also called plasmon modes have a finite dipolar moment and are therefore also accessible in infrared absorption experiments, the spin-density excitations do not have a dipolar moment and are only accessible in RILS. Under application of a magnetic field $B_\perp$ perpendicular to the QW plane, the individual subbands fan out in degenerate Landau levels separated by the cyclotron frequency $\hbar \omega_C = \frac{eB_\perp}{m^*}$ with the elementary charge *e* and the perpendicular applied magnetic field $B_\perp$ [54]. Each LL is further spin-split by the Zeeman energy $E_Z = g \mu_B B_{tot}$ depending on the material specific Landé *g*-factor, the Bohr magneton $\mu_B$ and the total magnetic field $B_{tot}$ [54].

In the following, we focus on the zero-field intersubband excitations (ISBE) determined by low-temperature RILS experiments. In Figure 2(b), typical RILS spectra of a 2DES hosted in a 31nm wide GaAs QW grown on the nonpolar (110) surface with transport density and mobility of $n = 1.0 \cdot 10^{11} cm^2$ and $\mu = 0.3 \cdot 10^6 cm^2/Vs$, respectively. In order to demonstrate the universality of the concepts, we purposely selected a sample with lower density and rather low mobility with the later given by less-ideal growth conditions for the nonpolar surface compared to the more commonly used polar (001) surface. The measurements are taken at 4.5K under resonant excitation for both, polarizing (blue traces) and depolarizing (red trace) scattering geometries. Three prominent RILS modes are observable. For both polarizations, a rather broad and weaker one centered at the expected energy spacing of the two lowest single-particle eigenstates of the QW $\Delta E_{01} = E_1 - E_0$ indicating the coexistence of spin-conserving and spin-flip transitions. The sharp mode in the depolarized spectra is a long-wavelength ISBE – spin density excitation (SDE). The SDE is red-shifted by exchange Coulomb interaction that is attractive for electrons [2]. For the long-wavelength plasmon-like ISBE - CDE observable in polarized scattering geometry direct and exchange Coulomb interaction terms cause a blue-shift [2]. The significant shift of the SDE mode with respect to the SPE mode reveal strong exchange interaction in the 2DES [2]. A schematic of the energy-momentum dispersion of the collective ISBE in the long-wavelength limit for $q \ll k_F$ with the Fermi wave-vector $k_F$ is inset in Figure 2(b). The CDE and SDE dispersion are rather independent from the in-plane momentum $q$. The single particle continuum is centered at a $\Delta E_{01}$ und homogeneously broadened with increasing $q$ by $2\hbar q v_F$, with the Fermi velocity $v_F$. An additional term that is vanishingly small for low momentum needs to be considered for larger momenta such that the dispersion is given by $\Delta E_{10} + \frac{\hbar^2}{m^*}\left(\frac{1}{2}q^2 \pm q k_F\right)$.

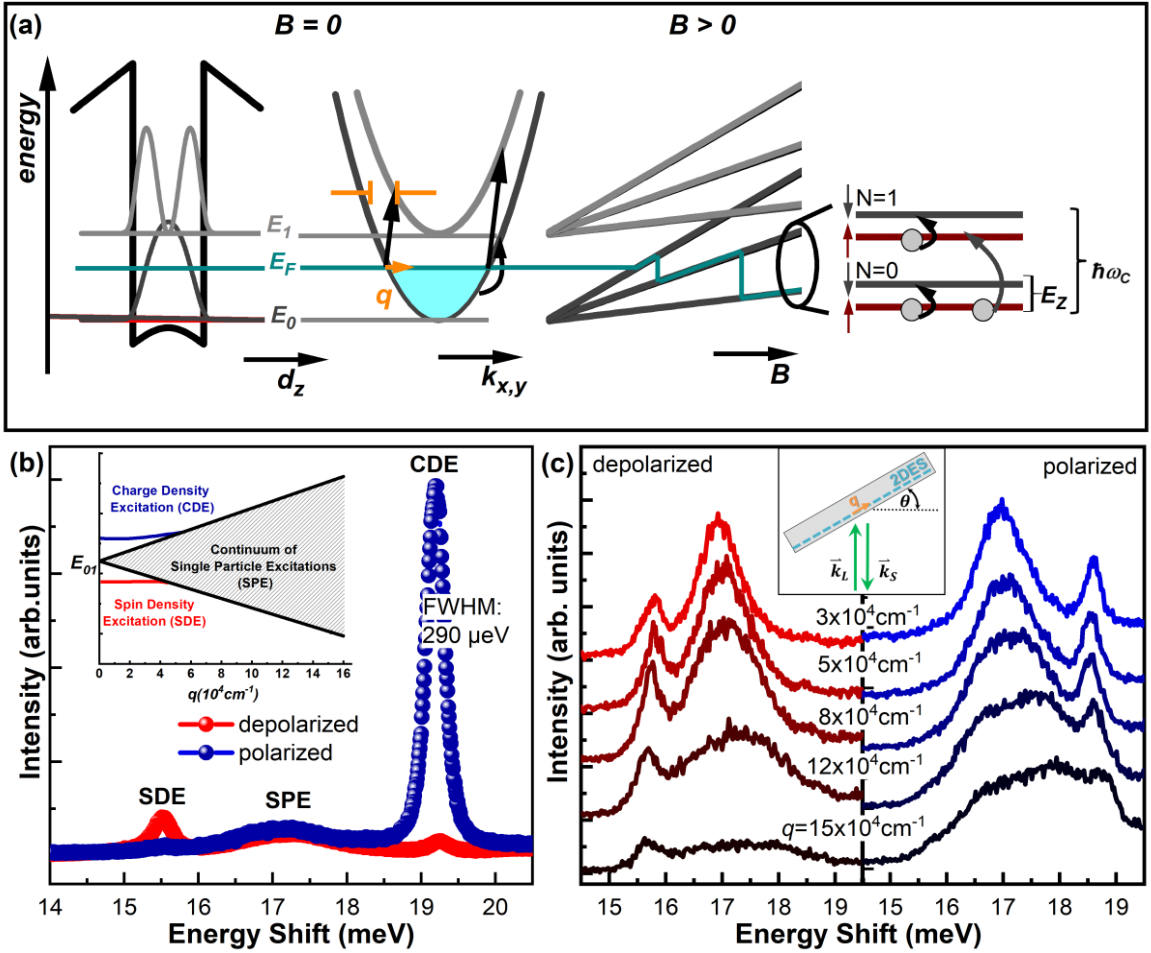

Figure 2: Simulated CB and first two eigen-states with energies $E_0$ and $E_1$ in the QW region together with the wave-function squared $|\psi|^2$ with $E_F$ and the first two subbands in k-space in the parabolic approximation with intra and intersubband excitations indicated with finite momentum transfer q. Scheme of the Landau level (LL) fan diagram for both subbands as a function of $B_\perp$. Next to it zoom-in into the two lowest LL of $E_0$. Intra-LL spin excitation across the Zeeman energy $E_Z$ and inter LL charge excitation across the cyclotron frequency $\hbar\omega_C$ are depicted. (b) RILS spectra under resonant condition taken at 4.5K on a single-side modulation-doped 31nm GaAs QW grown on the (110) surface with an electron density $n = 1.0 \cdot 10^{11} cm^{-2}$ and a mobility of $\mu = 0.3 \cdot 10^6 cm^2/Vs$. Linear polarized spectrum in linear co-polarized geometry is plotted in blue and depolarized spectrum in linear cross-polarized geometry is plotted as red trace. The so called single-particle excitation (SPE) band occurs in both scattering geometries and is a direct measure for the intersubband separation $\Delta E_{01}$. The spin-density excitation (SDE) is only excited in the depolarized geometry and is red-shifted with respect to the SPE energy, while the charge-density excitation (CDE) occurs selectively in the polarized geometry and is blue shifted with respect to SPE mode. A schematic of the dispersion of the collective intersubband excitations of a quasi-2D charge carrier system is shown in the inset. (c) Depolarized (left) and polarized (right) RILS spectra showing ISBE for different transferred momentum q realized by changing the angle between sample with respect to the incoming light as indicated by the scheme in the inset. The increasing width of the SPE band is in good agreement with the expected dispersion. The significant weakening and broadening of the SPE and CDE modes are caused by Landau damping. The inset indicates the back scattering geometry.

The dispersion is probed by momentum dependent RILS measurements on the identical sample summarized in Figure 2 (c) for depolarized spectra on the left and polarized spectra on the right. All measurements are done in back-scattering geometry with an in-situ mechanically rotatable cold finger such that the finite in-plane momentum transfer in backscattering geometry given by $q = |\vec{k}_L - \vec{k}_S| \approx (2\omega_L/c)\sin\theta$ can be modified in a certain range by changing the tilt angle $\theta$ between sample plane and the incident/ scattered light with momentum $\vec{k}_{L(S)}$ as indicated in the inset of Figure 2 (c). The spectra are excited by the linearly polarized light from a tunable Ti:sapphire laser with the frequency $\omega_L$ finely tuned close to the fundamental band gap energy such that the ISBE modes in RILS are resonantly enhanced. The scattered light is dispersed with a triple grating spectrometer and recorded by a CCD camera with a combined resolution of <20µeV. Suitable polarization optics are employed for polarized and depolarized scattering geometries.

Taking RILS spectra tilt angle for values of $\theta$ between 10° and 70°, the transferred in-plane momentum can be changed between $0.3 \cdot 10^5 cm^{-1} \leq q \leq 1.5 \cdot 10^5 cm^{-1}$. In this momentum range the SPE spectra are following the expected dispersion and broaden with increasing momenta for transitions with and without spin-flip [see Figure 2(c)]. With increasing momentum $q$, the SDE and CDE modes spectrally overlap with the broadened SPE continuum. In this regime, the line-widths of both plasmon-like modes, the SDE and CDE are increased and the mode intensities weakened due to Landau damping [2]. In comparison of the depolarizing and polarizing spectra, the effect of Landau damping is less pronounced for SDE in the investigated sample since the mode energy is further away from the $q\to 0$ value of the SPE mode. The origin of the SPE mode was long time debated [4],[7],[10],[55]. For historical reasons, this mode is called "single particle excitation" (SPE) because its energy in the long-wavelength limit is the subband spacing for vertical transitions. Das Sarma and Wang developed a stringent theory that the SPE mode is indeed a collective excitation of the interacting charge carrier system that does not appear in RILS because of disorder but is only accessible under extreme resonance conditions [5]. This explanation is corroborated by the finding that the SPE mode is better pronounced in samples with less disorder as can e.g. seen by comparing the resonantly enhanced RILS spectra on ISBE shown in Figures 2(b, c) on a lower mobility sample ($\mu = 0.3 \cdot 10^6 cm^2/Vs$) with the spectra shown in Figure 3(a) on a higher mobility sample with significantly reduced disorder ($\mu = 20 \cdot 10^6 cm^2/Vs$). The role of the resonance mechanisms has been further elaborated by Jusserand *et al.* demonstrating for 2DES in GaAs that the SPE is enhanced in RILS at resonances with electron-hole transitions at the Fermi wave vector $k_F$, and the plasmon-like modes at resonances with zone-center excitons [7]. We find that such moderate differences in the resonance conditions for the three ISBE modes are more distinct for samples with higher carrier densities and hence larger Fermi wave vector $k_F$ as expected.

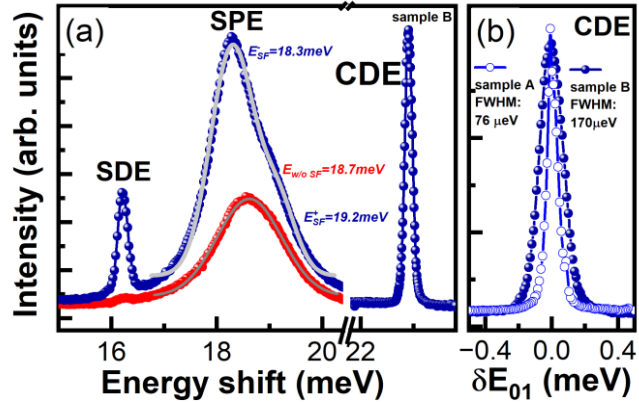

Figure 3: (a) Polarizing (blue) and depolarizing (red) RILS spectra on ISBE taken on an ultrahigh-mobility and high density 2D electron system confined in a 30nm GaAs QW. The SDE and CDE lines appears rather sharp while the SPE band is quite strong and appears split in the depolarizing spectrum with red-and blue-shifted contributions compared to the SPE in the polarized spectrum. The polarized SPE band is well described by single Gaussion function (solid line fit to the data), while a sum of two Gaussian function fit the depolarized SPE spectrum reasonably well. The likely origin for the double peak structure for the spin-flip transition is a spin-orbit coupling (SOI) induced splitting of parabolic bands resulting in two SPE continua. (b) CDE spectra of two 30nm GaAs QW samples on (001)-surface with similar transport characteristics of 2DES with similar density and mobility showing significant differences in the line-width of the mode demonstrating its sensitivity to probe the interacting electron systems. [T = 4.5K; B= 0T; resonance scattering conditions; transport characteristics of sample A: $n_t = 2.92 \cdot 10^{11} cm^2, \mu = 24 \cdot 10^6 cm^2/Vs$ and of sample B: $n_t = 3.2 \cdot 10^{11} cm^2, \mu = 20 \cdot 10^6 cm^2/Vs$, Fermi-energy and density from PL for sample A: $E_F = 12.2\ meV, n_{PL} = 3.2 \cdot 10^{11} cm^{-2}$ and for sample B: $E_F = 16.0\ meV, n_{PL} = 4.2 \cdot 10^{11} cm^{-2}$].

In addition to the better pronounced SPE mode in the RILS spectra from the higher quality sample shown in Figure 3(a), there is a significant difference in the lineshape of the depolarized (red trace) and polarized (blue trace) spectra. While the latter for the spin-conserving SPE can be well reproduced by a single Gaussian line profile (solid line overlaid the experimental data points), the SPE with spin-flip shows a double peak structure well described by a sum of two Gaussians [solid lines in Figure 3(a)]. We interpret the doublet structure in the SPE spectrum with spin-flip by a zero-field spin splitting $\Delta E(k_F)$ of the subbands due to spin-orbit interaction (SOI) effects caused by Dresselhaus and presumably Rashba type contributions with the latter due to a non-perfectly symmetric QW [56],[50]. Since the splitting of the subbands is expected to be identical for the first $E_0$ and second subband $E_1$ due to the parabolic CB, the SOI effects does not impact the spin-conserving SPE transitions. The value of the zero-field spin splitting $\Delta E(k_F)$ can by directly estimated from the splitting Δ of the SPE mode with spin-flip split for small $q \to 0$ assuming $\Delta = 2E(k_F)$. This interpretation is supported by the fact that the spin-conserving SPE band is symmetrically centered between the doublet of the spin-split SPE spectrum. In this way ISBE-SPE are suitable to determine and even quantify SOI induced splitting effects similar as reported for collective intraband excitations [33],[30],[31].

The larger charge carrier density in this sample increases the direct Coulomb interaction term and consequently the blue-shift of the CDE mode from the SPE mode is increased as well. It is reported in literature that the ISBE mode is rather sharp and nearly independent from inhomogeneous broadening [2]. Comparing the full width at half maximum (FWHM) of the CDE modes of the lower mobility sample shown in Figure 2(b) and the high-mobility sample shown in Figure 3(a) it appears that the FWHM is more than halved from 360µeV to 170µeV, while the mobility is enlarged by a factor of 67. While this comparison suggests a clear dependence between CDE linewidth and mobility, and hence purity of the 2DES, it is astonishing that the linewidth of a similar high mobility sample is further reduced by a factor of two from FWHM≈0170µeV for sample B to FWHM < 80µeV for sample (A) while the mobilities and transport densities differs only slightly with $\mu = 20 \cdot 10^6 cm^2/Vs$ and $n_t = 3.2 \cdot 10^{11} cm^2$ for sample B and $\mu = 24 \cdot 10^6 cm^2/Vs$ and $n_t = 2.9 \cdot 10^{11} cm^2$ for sample A, respectively. The densities determined optically from PL spectra constitute $n_{PL} = 4.2 \cdot 10^{11} cm^2$ and $n_{PL} = 3.2 \cdot 10^{11} cm^2$ for sample B and sample A, respectively, and show therefore a moderate increase compared to the transport values. This increase is presumably due to photo-activation of additional donor states. Magnetotransport investigations on sample B show evidence for a parallel conducting channel (Shubnikov-de Haas oscillations in the longitudinal magnetotransport measurements are no longer vanishing and increase with decreasing filling factor [57]) that is interpreted to be caused by high-doping in the Si δ-doping layer such that the CB shifts below $E_F$ in the doping layers [c.f. fig 1 (a)]. From optical experiments we can exclude that a second occupied subband is responsible for the parallel conductance, since the intersubband-spacing determined from RILS on the SPE mode is $E_{01}$ = 18.7meV and hence well above the Fermi energy of $E_F$ = 16 meV (for $n_{PL} = 4.2 \cdot 10^{11} \text{cm}^2$). In other words, the Fermi-energy is smaller than the intersubband-spacing. For $E_F < E_{01}$, a second subband cannot be occupied. For this reason, we do not expect that in this situation a parallel conducting channel negatively impacts the behavior of the interacting 2DES particularly for pure optical investigations. Even the opposite effect is assumed according to our previous experience. Free carriers in the δ-doping layer are expected to screen ionized donors effectively reducing the negative impact of ionized impurity scattering in the 2DES in the QW plane.

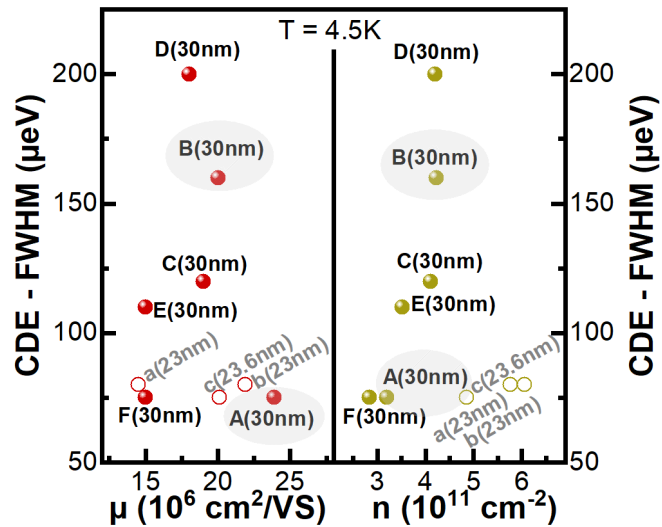

*Figure 4: Comparison of the line-widths (FWHM) of the CDE mode in dependence of mobility μ in (a) and density n in (b) for different 2DES confined in GaAs QWs on (001) GaAs. The samples have a QW width of either 30nm or 23nm for intermediate and high charge carrier densities, respectively. No clear relation between mobility μ and density n appear.*

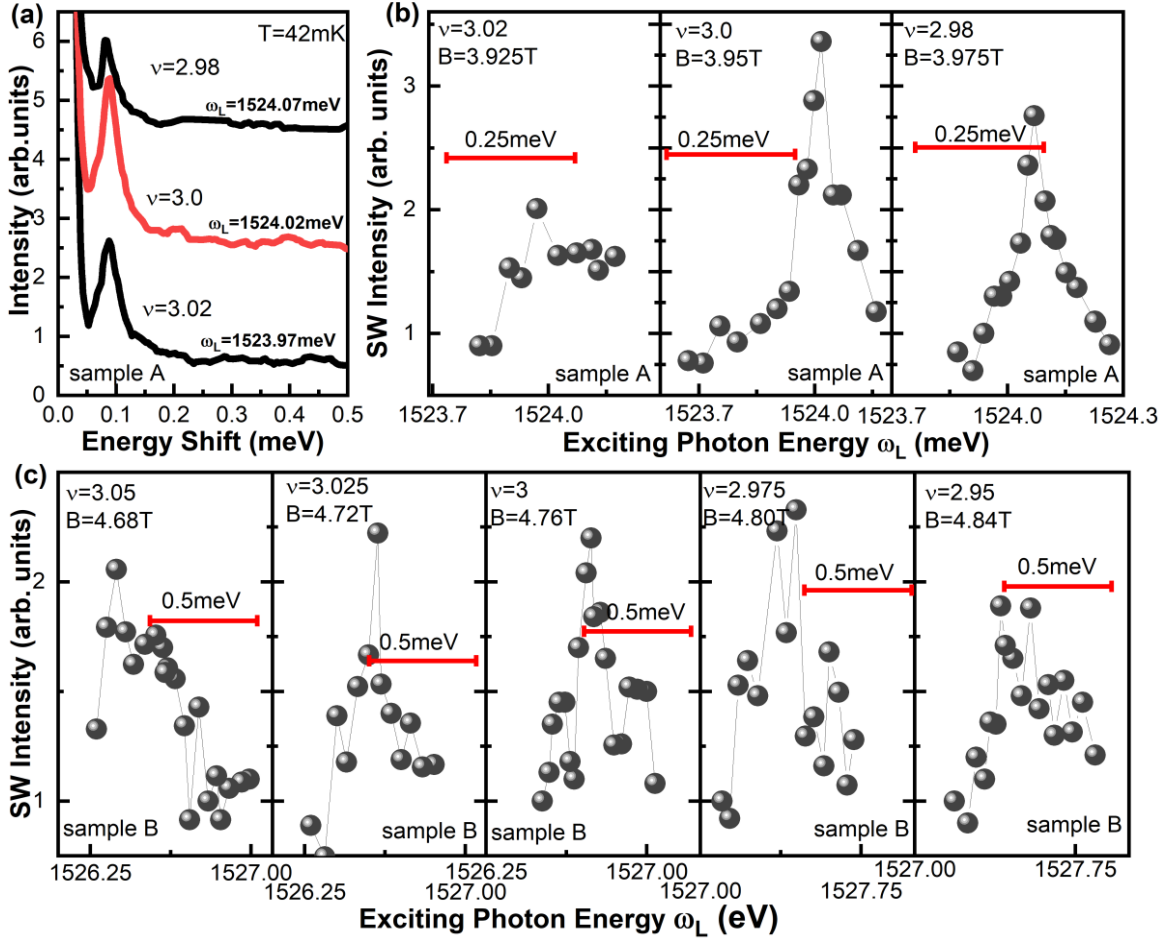

Figure 5: (a) Resonantly enhanced spin-wave (SW) modes around quantum Hall state $\nu = 3 \pm 0.02$ taken on sample A at a cold-finger temperature of about 42mK. The mode appears at $E_Z$ and the intensity is weakened for filling factors deviating from $\nu = 3$, the state in the second LL with the largest spin polarization allowing to identify this filling factor with high accuracy. (b) Resonance profiles for the SW modes at $\nu = 3$, $\nu = 3 \pm 0.02$ taken on sample A (resonant modes shown in (a)). The resonance profiles are rather sharp with a FWHM of < 0.08 meV and the fast decay of the SW intensity by slightly changing the magnetic field indicate an ultra-high quality 2DES with homogeneous density and ultra-low disorder resulting in sharp resonance profiles given by the (disorder) broadening of the underlying single-particle states. The solid lines are guides to the eye. (c) Resonance profiles for the SW modes at $\nu = 3$, $\nu = 3 \pm 0.025$, $\nu = 3 \pm 0.05$ taken on sample B. The resonance profile seems to consist of one main and two side resonances with a FWHM of the main line of < 0.11 meV. The intensities of the resonant enhanced SW mode are similar for $\nu = 3 \pm 0.025$ and hence less sensitive to the applied magnetic field indicating inhomogeneous carrier densities. The less distinct resonance profile indicates larger (disorder) broadening of single particle states with presumable multiple state close in energy. [T =42mK; tilt between sample normal and magnetic field about 20°; transport characteristics of sample A: $n = 2.92 \cdot 10^{11} cm^2, \mu = 24 \cdot 10^6 cm^2/Vs$ and of sample B: $n = 3.2 \cdot 10^{11} cm^2, \mu = 20 \cdot 10^6 cm^2/Vs$. Density from PL for sample A: $n_{PL} = 3.2 \cdot 10^{11} cm^{-2}$ and for sample B: $n_{PL} = 4.2 \cdot 10^{11} cm^{-2}$].

To further investigate the correlation between ISBE-CDE linewidth, transport mobility and density $n_{PL}$, several high quality and high-mobility 2DES samples with either 30nm wide QWs hosting densities between $3 - 4 \cdot 10^{11} cm^{-2}$ and for 23nm wide QWs hosting 2DES with densities ranging between $4.5 - 6 \cdot 10^{11} cm^{-2}$ are compared in Figure 4. The mobilities of the samples are very high with values between $15 \cdot 10^6 cm^2/Vs \leq \mu \leq 24 \cdot 10^6 cm^2/Vs$. The values for the FWHM of the CDE mode in RILS are determined under similar conditions as described above. In Figure 4, the FWHM values are plotted as a function of mobility μ and density $n_{PL}$. The values for the FWHM vary between 200μeV and 75μeV. For these high-mobility samples, we do not find clear correlation between FWHM, mobility and density. For this reason, we assume that the linewidth of the collective plasma-like CDE mode, that is connected to coherence [49],[58], might be a sensitive probe for the quality and purity of and interacting 2DES that cannot be distinguished from low-temperature zero-field transport experiments that probe Drude-transport parameters and hence are dominated by single-particle behavior.

In order to test this hypothesis, we investigate the evolution and enhancement profiles of a prominent collective spin-density mode at QHE filling factor $\nu = 3$ as sensitive probe of the quality for studies of correlated phases of interacting 2DES. The intra-LL spin-wave (SW) mode appears in the long-wavelength limit at the bare Zeeman energy due to Lamor's theorem. The intensity of the SW mode is sensitive to the spin-polarization in the second LL (N= 1). As visualized in Figure 5(a), the maximum of the SW intensity in RILS measurements precisely identifies filling factor $\nu = 3$ as a function of magnetic field [41]. The filling factor range $\nu = 3 \pm \delta$ at which the SW mode energy is maximal is thus an indication for fluctuations and spatial inhomogeneities in the charge carrier density within the probed spot that is in our case macroscopically large with 0.3mm x 25mm. The linewidth of the enhancement profile of a RILS mode is a measure for the disorder broadening of the single particle states involved in the resonant enhancement process, i.e. VB state and CB state which is here the N=1 LL. In Figure 5 (b, c), the resonance profiles in RILS from the SW mode around $\nu = 3 \pm \delta$ are summarized for sample (A) and sample (B), respectively. For sample A, the SW modes clearly diminish for minor changes in magnetic field and are already significantly reduced for $\nu = 3 \pm 0.02$. This fast disappearance of the SW mode with minor changes in magnetic field and hence nominal filling factor is a strong indication for a homogeneous density over the rather large investigated area of about 0.3mm x 25mm. The resonance profiles are quite sharp with a FWHM of < 80μeV similar as for the zero-field CDE mode indicating an ultra-high quality 2DES with homogeneous density and ultra-low disorder. For sample B, the resonance profile seems to consist of one main and two side resonances with a FWHM of the main line of > 110μeV that is even larger assuming the total FWHM including the side resonances. The intensities of the resonantly enhanced SW mode are similar for $3 - 0.025 \leq \nu \leq 3 + 0.025$ and hence less sensitive to the applied magnetic field indicating inhomogeneous carrier densities over the investigated sample area of again 0.3mm x 25mm. The less clear developed resonance profile for sample B indicates larger (disorder) broadening of single particle states with presumable multiple state close in energy.

The outcome of the RILS measurement on the SW mode in the quantum Hall regime underpins the initial conjecture that sample A is better suitable for the study of correlated electron phases in the (fractional) quantum Hall regime compared to sample B. We would like to emphasize that sample B is still of very high quality. For sample A, gapped mode in RILS for a series of FQHE states in the second LL (SLL) have been observed including those at the non-abelian quantum Hall state $\nu = 5/2$ as reported in literature [49],[37],[39],[41]. For sample B the weak modes of these fragile FQHE state in the SLL were not clearly observable. The magneto-RILS spectroscopy corroborates our initial claim that already the line-width of the ISB-CDE mode at rather high temperatures of 4.5K is a sensitive probe for the purity of a 2DES hosting heterostructure and the suitability of those structures for studying interaction-driven correlated quantum phases, particularly in the FQHE regime.

**Conclusion**

In summary we demonstrated that studying ISBE by RILS is a powerful method to characterize the quality and purity of interacting 2D electron systems confined in MBE-grown ultra-pure and low-disorder GaAs QWs. The so called SPE excitation that is despite its historical name a collective mode of the quasi-2D charge carrier system activated under extreme resonance conditions [5],[7], is more pronounced in high-quality less disordered heterostructures. Comparison of spectra taken in depolarizing and polarizing geometries provides access to zero-field spin-splitting of electronic bands due to Rashba- and Dresselhaus-type spin-orbit coupling effects [50]. We would like to mention that also collective intraband excitations accessed in RILS are sensitive to spin-split bands [33],[30],[31] and the methodology can also be applied to valance band states in GaAs heterostructures [6] and more generalized to mixed dipolar exciton-hole systems in tunnel-coupled double quantum wells [59] as well as to collective hole excitations between moiré-minibands in twisted WSe$_2$ bilayers [36]. Moreover, the line-width of the plasmon-like ISBE-CDE observable in polarized RILS measurements without external magnetic field has been introduced as highly sensitive probe for the impact of residual disorder and fluctuations in the charge carrier densities on the properties of the interacting charge carrier systems that are not detectable from zero-field transport characterization in ultra-high mobility 2DES in GaAs exceeding mobilities of $20 \cdot 10^6 cm^2/Vs$. This finding was corroborated by comparing the resonance profiles and appearance of the nominal filling factor range of the SW mode around quantum Hall filling factor $\nu = 3$. We would like to highlight that despite similar zero-field transport characteristics, for sample A with signature pointing towards lower disorder, robust gapped modes in RILS of the enigmatic and non-abelian FQHE state at $\nu = 5/2$ were reported [41]. However, no evidence for such modes was found in very similar experimental condition for the still ultra-high mobility 2DES of sample B with the slightly larger disorder according to results from magneto-RILS and zero-field RILS on CDE mode. These observations qualify RILS on ISBE as a fast and versatile method for the characterization of highest quality heterostructures hosting interacting low-dimensional electron systems at relaxed experimental conditions of a rather high-temperature of about 4.5K, without need of magnetic fields and electrical contacts. Since the methodology is suitable as local probe with diffraction limited lateral resolution, it can support further optimizing these structures serving as indispensable platform for the observation of novel quantum phases for fundamental studies [42],[43],[60] but also as test-bed structures to examine concepts in the quantum information processing section such as fault-tolerant quantum computation using non-abelian FQHE states [61],[62]. Since its establishment about 50 years ago and the seminal work of Aron Pinczuk and others, RILS on collective charge and spin density excitations of low-dimensional interacting electron system has been established as powerful method to analyze the fundamental properties of those systems and to explore aspect of correlated quantum phases, in particular the wave-vector dispersions of their low-lying neutral excitation spectra unique to each phase that are inaccessible with other methods.

**Acknowledgement**

With honor and deep gratitude, we dedicate this paper to Professor Aron Pinczuk, who pioneered the experiments and the understanding of collective neutral electronic excitations in GaAs based low-dimensional electron systems and enabled a fundamental understanding of novel quantum systems and exotic states of matter including (non-abelian) fractional quantum Hall liquids and phases in artificial graphene. We had the pleasure to work together and learn from Aron Pinczuk who actively contributed to the work presented in this manuscript – actually the optical experiments have been executed in his labs at Columbia university. It was a real joy collaborating with Aron Pinczuk and learn from and with his virtually endless curiosity and passion for neutral excitations in solids and new phases of matter. He will be always remembered for his dedication for physics combined with continued mentorship, positive attitude and sociability always connecting his impressive encyclopedic knowledge of physics with the human beings behind and their personalities and common paths.

The work was financial supported by the German Science Foundation (DFG) for financial support via Grants No. 443274199 (WU 637/7-2).